\documentclass[aps,prd,eqsecnum,12pt,showpacs,preprintnumbers,nofootinbib]{revtex4-2}
\usepackage{geometry}                
\usepackage[pdftex]{graphicx}
\usepackage{float}
\usepackage{amssymb,amsmath,amsopn,bm,dsfont,mathrsfs}

\usepackage[mathscr]{euscript}

\usepackage{silence}
\usepackage{enumitem}
\usepackage{txfonts}
\usepackage[compact,small]{titlesec}

\usepackage{amsbsy}

\IfFileExists{dsfont.sty}
	{\usepackage{dsfont}
         \let\mathbb=\mathds
         \newcommand{\id}{\mathds{1}}}
	{\typeout{Package dsfont.sty was not found, using alternative macros.}
         \let\mathds=\mathbb
         \newcommand{\id}{\mbox{1 \kern-.59em \textrm{l}}}}

\newcommand\del{\delta}
\newcommand\ka{\kappa}
\newcommand\la{\lambda}

\newcommand\m{\mu}
\newcommand\n{\nu}

\newcommand\p{\pi}
\newcommand\h{\eta}
\newcommand\s{\sigma}
\newcommand\f{\phi}
\newcommand\w{\omega}
\newcommand\ve{\varepsilon}

\newcommand\La{\Lambda}

\newcommand\D{\Delta}

\newcommand{\lag}{\langle}
\newcommand{\rag}{\rangle}

\newcommand{\cL}{{\cal L}}

\newcommand{\cO}{{\cal O}}
\newcommand{\cS}{{\cal S}}

\newcommand{\na}{\nabla}
\newcommand{\sdfrac}[2]{\mbox{\small$\displaystyle\frac{#1}{#2}$}}

\newcommand{\GN} {G_{\scriptstyle N}}
\newcommand{\rH}{\raise-5pt \hbox{$r_{\!_H}$}}
\newcommand{\rM}{r_{\!_M}}
\newcommand{\bea}{\begin{eqnarray}}
\newcommand{\eea}{\end{eqnarray}}
\newcommand{\be}{\begin{equation}}
\newcommand{\ee}{\end{equation}}
\newcommand{\bes}{\begin{subequations}}
\newcommand{\ees}{\end{subequations}}

\def\nbox#1#2{\vcenter{\hrule \hbox{\vrule height#2in
\kern#1in \vrule} \hrule}}
\def\sq{\,\raise0.8pt\hbox{$\nbox{.10}{.10}$}\,}
\def\sqb{\,\raise.5pt\hbox{$\overline{\nbox{.09}{.09}}$}\,}

\usepackage[T1]{fontenc} 
\usepackage[titletoc]{appendix}
\usepackage{setspace}

\allowdisplaybreaks
\usepackage[colorlinks,linkcolor=blue, pdfstartview=FitH]{hyperref}
\hypersetup{linkcolor=blue,citecolor=blue,filecolor=black,urlcolor=blue}

\begin{document}
\pagestyle{empty}

\title{Gravitational Condensate Stars: An Alternative to Black Holes}

\author{Emil Mottola}
\email{mottola.emil@gmail.com, emottola@unm.edu}
\affiliation{Department of Physics and Astronomy, University of New Mexico\\
Albuquerque NM 87131}

\author{Pawel O. Mazur}
\email{pomazur@mailbox.sc.edu}
\affiliation{Department~of Physics and Astronomy, University~of South Carolina\\ Columbia, SC 29208, USA}

\begin{abstract}
A new final endpoint of complete gravitational collapse is proposed. By extending the concept of Bose--Einstein 
condensation to gravitational systems, a static, spherically symmetric solution to Einstein's equations is obtained,
characterized by an interior de Sitter region of $p \!=\! -\rho$ gravitational vacuum condensate and an exterior 
Schwarzschild geometry of arbitrary total mass $M$. These are separated by a phase boundary with a small but 
finite thickness $\ell$, replacing both the Schwarzschild and de Sitter classical horizons. The resulting collapsed
cold, compact object has no singularities, no event horizons, and a globally defined Killing time. Its entropy is 
maximized under small fluctuations and is given by the standard hydrodynamic entropy of the thin shell, which 
is of order $k_B\ell Mc/\hbar$, instead of the Bekenstein--Hawking entropy, $S_{BH}= 4\pi k_B \GN M^2/\hbar c$. 
Unlike black holes, a collapsed star of this kind is consistent with quantum theory, thermodynamically stable, 
and suffers from no information paradox.
\end{abstract} 

\maketitle
\vfil
\eject


\pagenumbering{arabic}
\pagestyle{plain}
\vfil\break

\setcounter{page}{1}

\section{Introduction}
\label{Sec:Intro}

The vacuum Einstein equations of classical general relativity (GR) possess a well-known solution for an isolated mass 
$M$, with~the static, spherically symmetric line element
\vspace{-2mm}
\be
ds^2 = -f(r)\, dt^2 + \frac{dr^2}{h(r)} + r^2 \big( d\theta^2 + \sin^2\theta\,d\f^2\big)
\label{sphsta}
\vspace{-2mm}
\ee
where the functions $f(r)$ and $h(r)$ in this case are equal and given by
\vspace{-2mm}
\be
f(r) = h(r) = 1 - \frac{2\GN M}{r}  = 1 -\frac{\rM\!}{\!r}
\label{Sch}
\vspace{-2mm}
\ee
in units where $c=1$.  The~dynamical singularity of the Schwarzschild metric (\ref{sphsta}) at $r=0$ 
with its infinite tidal forces clearly signals a breakdown of the vacuum Einstein equations.
The kinematical singularity at the Schwarzschild radius $\rM \!\equiv\! 2\GN M$ is of a different sort, 
corresponding to an infinite blue shift of the frequency of an infalling light wave with respect to its 
frequency far from the black hole (BH). Since the local curvature tensor is finite at $r=\rM$, the~singularity 
of the metric (\ref{sphsta})--(\ref{Sch}) there can be removed by a suitable (and singular) change of 
coordinates in the classical theory. A~classical point test particle freely falling through the event horizon 
is said to experience nothing special at $r=\rM$. Whether or not a true event horizon of this kind
where light itself becomes trapped can be realized in a physical collapse process remains open to~question.

This question becomes much more acute in quantum theory. For~when $\hbar \neq 0$, a~photon of finite asymptotic 
frequency $\w$ (even if arbitrarily small) acquires a local energy $E=\hbar \w \,[f(r)]^{-\frac{1}{2}}$, which 
diverges at $r\!=\!\rM$. Since the effective coupling in gravity is $(\GN/\hbar)^\frac{1}{2} E$, proportional to energy,
the kinematical singularity at $\rM$ may be responsible for strong gravitational interactions between 
elementary quanta as their energy approaches the Planck energy $M_{Pl}\!=\! (\hbar/\GN)^\frac{1}{2}$, by
which point it can no longer be taken for granted that quantum effects on the classical geometry can be 
safely~neglected. 

In the semi-classical approximation, when a massless field, such as that of the photon, is quantized in the fixed 
Schwarzschild background, with~certain boundary conditions corresponding to a regular future horizon, one finds 
that the BH radiates these quanta with a thermal spectrum at the asymptotic Hawking temperature,
$T_H = {\hbar/8\pi k_B \GN M}$ \cite{Hawking:1974}. It is usually assumed that the backreaction effect of this 
radiation on the classical geometry must be quite small. However, detailed calculations of the energy-momentum of 
the radiation show that its $\lag T^t_{\ t} \rag$ and $\lag T^r_{\ r} \rag$ components have an $f^{-2}$ infinite 
blue shift factor at the horizon which divergences are arranged to exactly cancel in free-falling coordinates~\cite{BirDav,ChrFul:1977}. 
Anything other than this exact cancellation of the separately diverging energy density and pressure in the semi-classical 
Einstein equations would significantly change the geometry near $r=\rM$ from the classical Schwarzschild solution 
(\ref{Sch}). The~wavelengths contributing to these quantum stresses are of order $\rM$ and, hence, quite non-local 
on the scale of the BH. Unlike the classical kinematic singularity in (\ref{Sch}), such non-local semi-classical 
backreaction effects near $\rM$ depending on the quantum state of the field theory, cannot be removed by a 
local coordinate~transformation. 

Furthermore, energy conservation plus a thermal radiation spectrum imply that a BH has an enormous 
entropy, $S_{BH}\simeq 10^{77} k_{_B}(M/M_{\odot})^2$ \cite{Beken:1973}, far in excess of a typical 
stellar progenitor. The~application of thermodynamic arguments to BHs is itself put into doubt by the
the inverse dependence of $T_{_H}$ on $M$ implying that a BH in thermal equilibrium with its own Hawking 
radiation has negative specific heat and, therefore, is unstable to thermodynamic fluctuations~\cite{HawkSpecHeat:1976}.
On the other hand, requiring that basic thermodynamic principles apply to self-gravitating systems as well
implies that their heat capacity must be proportional to their energy fluctuations, $\propto \lag (\D E)^2 \rag$,
and hence must be positive.  The~assumption of a thermal mixed state of Hawking radiation emerging from a BH
also leads to an `information paradox' so severe that resolving it has been conjectured to require an alteration in 
the principles of relativity, or~quantum mechanics, or~both. 

In light of the challenges BHs pose to quantum theory, and~in lieu of revision to otherwise well-established
fundamental laws of physics, it is reasonable to examine alternatives to the strictly classical view of the event horizon 
as a harmless kinematic singularity, when $\hbar \neq 0$ and the quantum wavelike properties of matter are taken into 
account. In~earlier investigations which attempted to include the backreaction of the Hawking radiation in a self-consistent 
way, the~entropy arises entirely from the radiation fluid~\cite{ZurekPage:1984,tHooft:1998}. In~fact,
$S = 4\,[(\ka + 1)/(7 \ka + 1)]\,S_{BH}$, for~a fluid with the equation of state, $p =\ka \rho$, becoming equal to the 
Bekenstein--Hawking entropy $S_{BH}$ for $\ka = 1$. Despite this suggestive feature, these fluid models have 
huge (Planckian) energy densities near $\rM$ and a negative mass singularity at $r=0$, so that the Einstein equations are 
not reliable in either~region. 

A quite different proposal for incorporating quantum effects has been made in~\cite{ChaplineHCL:2001},
{\it viz.} that the horizon becomes instead a critical surface of a phase transition in the quantum theory, 
supported by an interior region with equation of state,  $p = - \rho < 0$. Such a vacuum equation of state, first proposed 
by Gliner~\cite{Gliner:1966} for the endpoint of gravitational collapse, is equivalent to a positive cosmological term
in Einstein's equations, and~does not satisfy the energy condition $\rho + 3p \ge 0$ needed to prove the most general
form of the classical singularity theorems~\cite{HawkEllis:1973}. 

In this paper~\cite{gravastar:2001}, we show that an explicit static solution of Einstein's equations taking quantum 
considerations into account exists, with~the critical surface of ref.~\cite{ChaplineHCL:2001} replaced by a thin shell 
of ultra-relativistic fluid of soft quanta obeying $\rho = p$. Such a solution, lacking a singularity and an horizon is 
significant because it provides a stable alternative to BHs as the endpoint of gravitational collapse, with~potentially
different observational~signatures.

The principal assumption required for this solution to exist is that gravity undergoes a vacuum rearrangement 
phase transition in the vicinity of $r=\rM$,  in~which the vacuum energy density changes abruptly. Since a spatially 
homogeneous Bose-Einstein condensate (BEC) couples to Einstein's equations in exactly the same way as an effective 
cosmological term $\La_{\rm eff}$, with~equation of state $p= -\rho$, the~existence of the interior region 
requires that general considerations of low temperature quantum BEC phase transitions can be extended 
to gravitation. It has been suggested that a phase transition could be induced by the equation of state of
the compressed matter attaining the most extreme one allowed by causality, namely $p = +\rho$. The~effective 
theory incorporating the low energy effects of quantum anomalies that could give rise to the interior $p=-\rho$
Gravitational Bose--Einstein condensate (GBEC) phase has been presented elsewhere~\cite{MazEMWeyl:2001,EMVau:2006,EMZak:2010}.
In this paper, we forego any discussion of the details of the quantum phase transition and present only the solution 
of Einstein's equations with the specified equations of state of the de Sitter (dS) interior and phase boundary layer, which is
the model proposed in the original arXiv paper~\cite{gravastar:2001}. Developments since that original article are discussed
in the Appendix under seven subheadings.

\section{Solution of Einstein Equations for Static, Spherical~Symmetry} 
\label{Sec:Soln}

The general form of the stress-energy tensor in the static, spherically symmetric geometry of (\ref{sphsta}) is
\vspace{-3mm}
\be
T_\m^{ \ \,\n} = \left(\begin{array}{cccl}\!\! -\rho\ & 0\ & 0\ & 0\\
0 & p & 0& 0\\ 0 & 0 & p_\perp& 0\\ 0 & 0 & 0 & p_\perp \end{array}\right) 
\label{Tgen}
\ee
so that the Einstein equations in the static spherical coordinates of (\ref{sphsta}) are 
\vspace{-2mm}
\begin{subequations}
\begin{align}
-G_{\ t}^t &= \sdfrac{1}{r^2} \sdfrac{d}{dr}\,\big[r\left(1 - h\right)\big] = -8\p \GN T_{\ t}^t = 8\p \GN \,\rho\,,\\[1ex]
G_{\ r}^r &= \sdfrac{h}{r f}\sdfrac{d f}{dr}  + \sdfrac{1}{r^2} \,\big(h -1\big) = 8\p \GN T_{\ r}^r = 8\p \GN\, p
\vspace{-2mm}
\end{align}
\label{Eins}\end{subequations} 
together with the conservation equation
\vspace{-2mm}
\be
\na\!_\la\, T^\la_{\ r} = \frac{d p}{dr} + \frac{\rho + p}{2f} \,\frac{d f}{dr} + \frac{2}{r} \,(p-p_\perp) = 0
\label{cons}
\vspace{-2mm}
\ee
which ensures that the other components of Einstein's equations are satisfied. In~(\ref{cons}) the transverse 
pressure $p_\perp \!\equiv\! T^{\theta}_{\ \theta} \!=\! T^{\f}_{\ \f}$ is allowed to be different from the radial pressure 
$p \equiv T^r_{\ r}$. For~a perfect fluid $p_\perp \!=\! p$ and the last term of (\ref{cons}) 
vanishes. In~that case, (\ref{Eins})--(\ref{cons}) are three first order equations for the four functions, $f, h, \rho$, and~$p$, 
which become closed when an equation of state for the fluid relating $p$ and $\rho$ is~specified.

Because of the considerations in the Introduction, as~a first phenomenological model we allow for three different regions
with the three different equations of state
\vspace{-4mm}
\be
\begin{array}{clcl}
{\rm I.}\ & {\rm de\ Sitter\ Interior:}\ & 0 \le r < r_1\,,\ &\rho = - p \,,\\
{\rm II.}\ & {\rm Thin\ Shell:}\ & r_1 < r < r_2\,,\ &\rho = + p\,,\\
{\rm III.}\ & {\rm Schwarzschild\ Exterior:}\ & r_2 < r\,,\ &\rho = p = 0\,.
\end{array}
\vspace{-3mm}
\ee

At the interfaces $r\!=\!r_1$ and $r\!=\!r_2$, the~metric functions $f$ and $h$ are required to be 
continuous, although~the first derivatives of $f$, $h$ and $p$ must be discontinuous from the first 
order equations~(\ref{Eins}) and (\ref{cons}). In the interior region $\rho = - p$ is a constant from (\ref{cons}).
Let us call this constant $\rho_{_V} = 3H^2/8\p \GN$. 
If we require that the origin is free of any mass singularity then the interior is determined to be a region of dS 
spacetime in static coordinates, {\it i.e.}
\vspace{-3mm}
\be
{\rm I.}\qquad f(r) = C\,h(r) = C\,\big(1 - H^2\,r^2\big)\,,\quad 0 \le r \le r_1
\label{deS}
\vspace{-3mm}
\ee
where $C$ and $H$ are constants, which at this point are~arbitrary.

The unique solution in the exterior vacuum region which approaches flat space as $r \to \infty$ is a region of 
Schwarzschild spacetime (\ref{Sch}), {\it viz.}
\vspace{-3mm}
\be
{\rm III.} \qquad f(r) = h(r) = 1 - \sdfrac{2 \GN M}{r} = 1-\sdfrac{\rM}{r}\,,\qquad  r_2 \le r
\vspace{-3mm}
\ee
where the mass $M$ can take on any (positive) value.

The only non-vacuum region is region II. Defining the dimensionless variable $w$ by 
\vspace{-4mm}
\be
w \equiv 8\p \GN r^2 p
\label{wdef}
\ee 
equations~(\ref{Eins}) and (\ref{cons}) with $\rho = p$ may be recast in the {form} 
\vspace{-2mm}
\begin{subequations}
\begin{align}
\frac{dr}{r} &= \frac{dh}{1-w-h}\,, \label{ueqa}\\[0.9ex]
\frac{dh}{h} &= -\frac{1-w-h}{1 + w - 3h}\, \sdfrac{dw}{w}\,.
\label{ueqb}
\end{align}
\label{ueq}\end{subequations} 
together with $p f \propto wf/r^2$ a constant. The~first equation ~(\ref{ueqa}) is equivalent to the definition 
of the (rescaled) Tolman--Misner--Sharp mass function $\m(r)=2\GN m(r)$, with~$h = 1 - \m/r$ and 
$d\m(r) = 8\p \GN\, \rho r^2\, dr = w\, dr$ within the shell. The~second equation~(\ref{ueqb}) can be solved only numerically 
in general. However, it is possible to obtain an analytic solution in the thin shell limit, $0 < h \ll 1$, for~in this limit 
we can set $h$ to zero on the right side of (\ref{ueqb}) to leading order, and~integrate it immediately to obtain
\be
h \equiv 1- \frac{\m}{r} \simeq  \ve\ \frac{(1 + w)^2}{w \ }\ll 1
\label{hshell}
\ee
in region II, where $\ve$ is an integration constant. Because~of the condition $h \ll 1$ we require 
$\ve \ll 1$, if~$w$ is of order unity. Making use of equations~(\ref{ueq}) and (\ref{hshell}) we have
\vspace{-2mm}
\be
\frac{dr}{r} \simeq - \ve\, dw\, \frac{(1 + w)}{w^2}
\label{req}
\vspace{-2mm}
\ee
so that because $\ve \ll 1$ the radius $r$ hardly changes within region II, and~$dr$ is of order $\ve \,dw$. 
The final unknown function $f$ is given by (\ref{cons}) to be $f = (r/r_1)^2 (w_1/w) f(r_1) \simeq  (w_1/w) f(r_1)$ to
leading order in $\ve$ for $\ve \ll 1$.

Now requiring continuity of the metric coefficients $f$ and $h$ at both $r_1$ and $r_2$ gives the~conditions
\vspace{-2mm}
\begin{subequations}
\begin{align}
f(r_1) &= C h(r_1) = C (1-H^2 r_1^2) \simeq C \ve \,\frac{(1+ w_1)^2\!\!}{w_1} \label{match1}\\ 
f(r_2) & = h(r_2) = 1 - \frac{2GM}{r_2} \simeq  \ve \,\frac{(1+ w_2)^2\!\!}{w_2}
\vspace{-2mm}
\end{align}
\label{match}\end{subequations}
which together with the solution for $f$ evaluated at $r=r_2,w=w_2$ gives
\vspace{-2mm}
\be
C(1+w_1)^2 = (1 + w_2)^2
\label{Cw1w2}
\vspace{-3mm}
\ee
and three independent relations among the eight integration constants $(r_1, r_2, w_1, w_2, H, M, C, \ve)$. 
Assuming that $(r_1, r_2, w_1, w_2, H, M, C)$ all remain finite as $\ve\to 0$, i.e.~they are all of order $\ve^0$, 
we obtain from (\ref{req}) and (\ref{match}) that
\vspace{-4mm}
\begin{subequations}
\begin{align}
&r_1 = \frac{1}{H} \left[ 1 - \ve \, \frac{(1+w_1)^2\!\!}{2w_1}\ \right] \\ 
&r_2 =  \rM \left[ 1 + \ve\, \frac{(1+ w_2)^2\!\!}{w_2}\ \right]
\vspace{-2mm}
\end{align}
\label{r1r2C}\end{subequations}
for two of the three relations, so that $r_1\! \to\!  \rH =H^{-1}$ and $r_2\!  \to \! \rM$ with $r_2-r_1=\D r$ of order $\ve$, 
and $\rH \simeq \rM$ to leading order in $\ve$. Thus, the~boundary layer {\rm II} straddles the location 
of the classical Schwarzschild and dS horizons, and~$r_1 \to r_2$ coincide at $\rH = \rM$, becoming no
longer independent in the limit $\ve \to 0$. Since the mass $M$ is a free parameter there remain three 
undetermined integration constants $C, w_1 ,w_2$ which satisfy the one relation (\ref{Cw1w2}) if $\ve \neq 0$, 
and lastly $0<\ve \ll 1$ itself.

The important feature of this solution is that for any $\ve>0$ both $f$ and $h$ are of order $\ve$ but nowhere 
vanishing.  Hence there is no event horizon, and~$t$ is a global Killing time. A~photon experiences 
a very large, ${\cal O}(\ve^{-\frac{1}{2}})$ but finite blue shift in falling into the shell from~infinity.

The proper thickness of the shell in the metric (\ref{sphsta}) is
\be
\ell = \int_{r_1}^{r_2}\frac{dr}{\!\!\!\sqrt{h}} = \rM\! \sqrt{\ve}\int_{w_2}^{w_1}\! dw\, w^{-\frac{3}{2}}
= 2 \rM\! \sqrt{\ve}\,\Big(w_2^{-\frac{1}{2}} - w_1^{-\frac{1}{2}}\Big)
\ee
to leading order in $\ve$, and, hence $\ell$ is ${\cal O}(\ve^{\frac{1}{2}})$ and small compared to $\rM$. 

The magnitude of $\ve$ and hence of $\ell$ can be fixed only by consideration of the quantum effects that give rise 
to the phase transition boundary layer. A~subsequent analysis of the stress tensor of the conformal anomaly~\cite{EMVau:2006,EMZak:2010}, 
shows that these quantum vacuum polarization effects become significant 
when $r_2-\rM$ is of order of the Planck length $L_{\rm Pl} = (\hbar \GN)^{\frac{1}{2}} \simeq 1.6 \times 10^{-33}$ cm,
so that $\ve \sim L_{\rm Pl}/\rM$, and~$\ell \sim \sqrt{L_{\rm Pl}\rM} \gg L_{\rm Pl}$, making a a semi-classical
mean field treatment of the boundary layer~feasible.

The entropy of the thin shell is obtained from the equation of state, $p = \rho = (a^2/8\p \GN) (k_B T/\hbar)^2$, 
where we have introduced $\GN$ for dimensional reasons so that $a^2$ is a dimensionless constant. 
By the standard thermodynamic Gibbs relation, $T s = p + \rho$ for a relativistic fluid with zero chemical potential, and, 
hence, the~local specific entropy density~is 
\be 
s(r) = \frac{a^2k_B^2 T(r)}{4\p \hbar^2 \GN} =\frac{ak_B}{\hbar} \left(\frac{p}{2\p \GN}\right)^{\frac{1}{2}}
= \frac{ak_B}{4\p \hbar \GN r} \,w^{\frac{1}{2}}
\label{localent}
\ee
for local temperature $T(r)$. The~entropy of the fluid within the shell is thus
\vspace{-3mm}
\be
\hspace{-3mm}S =4\p\! \int_{r_1}^{r_2} \frac{s\, r^2\,dr}{\!\!\!\sqrt{h}\ }
= \frac{ak_{_B}\rM^2}{\hbar \GN} \sqrt{\ve} \,\ln \Big(\frac{w_1}{w_2}\Big)
\sim  a\, k_B \frac{M\ell}{\hbar}
\label{entsh}
\vspace{-2mm}
\ee
and of order $k_B M \ell/\hbar$ to leading order in $\ve$, assuming $a,w_1,w_2$ are $\cO(1)$. Since the interior region I has $\rho_{_V} = -p_{_V}$, 
$(T s)_{_V} = p_{_V} + \rho_{_V} $ vanishes there. This is in accord with a~GBEC having equation of state $\rho_{_V} = -p_{_V}$ being a single coherent
macroscopic quantum state with zero entropy. Thus, the~entropy of the entire compact quasi-black hole (QBH) is given by the entropy of the
shell alone. By~(\ref{entsh}) this is of order $k_B (\rM/L_{\rm Pl})^{\frac{3}{2}}$ for $\ell \!\sim \!\sqrt{L_{\rm Pl}\rM}$,
or $S\!\sim\! \sqrt{\ve}\, S_{BH} \ll S_{BH}$, far smaller than the Bekenstein--Hawking entropy. Its $M^{3/2}$ scaling, 
furthermore, makes it comparable to the entropy of typical stellar progenitors of mass $M$, in~the range of $10^{57} k_B$ 
to $10^{59} k_B$ for a solar mass and $M_\odot/m_N \sim 10^{57}$ nucleons. Thus there is no information paradox arising from
an enormous entropy unaccountably associated with a BH horizon, if~the horizon is replaced by a thin boundary layer of this~kind.

Since $w$ is of order unity in the shell, the~{\it local} temperature of the fluid within the shell is of order 
$T_H  \sim \hbar/k_B \rM$, so that the typical quanta are {\it soft} with wavelengths of order $\rM$, and~there
is no transplanckian problem.  Because~of the global timelike Killing field $t$ and absence of either an event horizon 
or an interior singularity, there is no loss of unitarity or conflict with quantum theory.  As~a static solution,
neither the interior nor the shell emit Hawking radiation. A~gravitational condensate star is both cold and dark, 
and, hence, in~its external geometry and its appearance to distant observers indistinguishable from a~BH.

The cold radiation fluid in the shell is confined to region II by the surface tensions at the timelike interfaces 
$r_1$ and $r_2$. These arise from the pressure discontinuities, $\D p_1 \simeq H^2\, (3 + w_1)/8\pi G$ and
$\D p_2 \simeq -w_2/ 32\pi G^3M^2$, and~are calculable by the Lanczos--Israel junction conditions
~\cite{Israel:1966a,Israel:1967,BerKuzTka:1987}. The~non-zero angular components of the surface tension 
are \footnote{The sign conventions in~\cite{gravastar:2001,MazEMPNAS:2004} are such that $\s_{1,2}$ there are the 
{\it negative} of the surface stress tensors $\cS^{\ \theta}_\theta =\cS^{\ \f}_\f$ properly defined here. 
Equations~(C5) and (C7) of~\cite{MazEM:2015} also have an overall sign change from the Lanczos--Israel formula 
(C5) for $\cS^{\ b}_a$, such that $\h,\s$ of (C7) in~\cite{MazEM:2015} have the same values as $\h,\s$ 
in~\cite{gravastar:2001,MazEMPNAS:2004}.}
\vspace{-3mm}
\begin{subequations}
\begin{align}
&\cS^{\ \theta}_{\theta}\big\vert_{r=r_1} = \cS^{\ \f}_{\f}\big\vert_{r=r_1}\equiv -\s_1= \frac{1}{32\p G^2M}\frac{(3 + w_1)}{(1 + w_1)}\left(\frac{w_1}{\ve}\right)^\frac{1}{2}\\[0.8ex]
&\cS^{\ \theta}_{\theta}\big\vert_{r=r_2}  = \cS^{\ \f}_{\f}\big\vert_{r=r_2}\equiv -\s_2 =-\frac{1}{32\p G^2M} \frac{w_2}{(1 + w_2)}\left(\frac{w_2}{\ve}\right)^\frac{1}{2}
\vspace{-5mm}
\end{align}
\label{surf}\end{subequations}
to leading order in $\ve$, at~$r_1$ and $r_2$, respectively. The~signs correspond to the inner surface at $r_1$ 
exerting an outward force and the outer surface at $r_2$ exerting an inward force, i.e.~both surface 
tensions exert a confining pressure on the shell region II. Clearly these large transverse surface tensions
violate the perfect fluid ansatz at the interfacial boundaries. Nevertheless, since 
$\ve^{-\frac{1}{2}} \sim (M/M_{\rm pl})^\frac{1}{2}$, the~surface tensions (\ref{surf}) are of order
$M^{-\frac{1}{2}}$ and far from Planckian, so that the matching of the metric at the phase interfaces 
$r_1$ and $r_2$, analogous to that across stationary shocks in hydrodynamics, should be reliable. 
The time component of the surface stress tensor at $r_1$ and $r_2$ vanishes and makes no
contribution to the Tolman--Misner--Sharp mass function $\m(r)=2Gm(r)$ at either of the two~interfaces. 

The Misner--Sharp energy within the shell 
\vspace{-3mm}
\be
E_{\rm II} = 4\p\! \int_{r_1}^{r_2}\!\rho\,r^2 dr = \ve M\! \int_{w_2}^{w_1}\frac{dw}{w}\,(1 + w)
= \ve M\, \bigg[\!\ln \Big(\sdfrac{w_1}{w_2}\Big) + w_1 - w_2\bigg]
\vspace{-3mm}
\ee
to leading order in $\ve$, is of order $M_{\rm Pl}$ and also extremely small. Hence essentially all the 
Misner--Sharp mass of the object comes from the energy density of the vacuum condensate in the interior, even though
the shell is responsible for all of its~entropy. 

\section{Stability} 
\label{Sec:Stability}

In order to be a physically realizable endpoint of gravitational collapse, any QBH candidate must be stable~\cite{Mazur:1996}. 
Since only  region II is non-vacuum, with~a `normal' fluid and a positive heat capacity, it is clear that the solution is
thermodynamically stable. The~most direct way to demonstrate this stability is to work in the microcanonical ensemble 
with fixed total $M$, and~show that the entropy functional
\vspace{-4mm}
\be
S=\frac {a k_{_B}}{\hbar \GN}\int_{r_1}^{r_2}r \,dr\, \left(\frac{d\m}{dr}\right)^{\!\frac{1}{2}}
\left(1 - \frac{\m(r)}{r}\right)^{\!-\frac{1}{2}}
\label{entropyfn}
\ee
for the $p=\rho$ fluid in region II is maximized under all variations of $\m(r)$ with the endpoints $(r_1, r_2)$, equivalently $(w_1, w_2)$ fixed.

The first variation of this functional with the endpoints $r_1$ and $r_2$ fixed vanishes, i.e.~$\del S =0$ 
by the Einstein equation~(\ref{Eins}) for a static, spherically symmetric star. Thus, any solution of equations~(\ref{Eins}) and (\ref{cons}) 
is guaranteed to be an extremum of $S$ \cite{Cocke:1965}. This is also consistent with regarding Einstein's equations 
as a form of hydrodynamics,  strictly valid only for the long wavelength, gapless excitations in gravity. 
The second variation of (\ref{entropyfn}) is 
\vspace{-3mm}
\be
\del^2 S=\frac{a k_B}{4\hbar \GN}\int_{r_1}^{r_2}\!r\,dr\, \left(\frac{d\m}{dr}\right)^{\!-\frac{3}{2}}
h^{\!-\frac{1}{2}}\left\{-\left[\frac{d(\del\m)}{dr}\right]^2 + \frac{(\del\m)^2}{r^2h^2}
\frac{d\m}{dr}\left(1+ \frac{d\m}{dr}\right) \right\}
\label{varent}
\vspace{-2mm}
\ee
when evaluated on the solution. Associated with this quadratic form in $\del\m$ is a second order linear differential 
operator $\cL$ of the Sturm--Liouville type, {\it viz.}
\vspace{-3mm}
\be
\cL \,\chi = \frac{d}{dr} \left\{r\left(\frac{d\m}{dr}\right)^{\!-\frac{3}{2}}\!
h^{\!-\frac{1}{2}} \frac{d\chi}{dr}\right\}+ \frac{h^{-\frac{5}{2}}}{r} 
\left(\frac{d\m}{dr}\right)^\frac{1}{2}\left(1+ \frac{d\m}{dr}\right)\, \chi \,.
\label{stli}
\vspace{-2mm}
\ee
This operator possesses two solutions satisfying ${\cL}\chi_{_0} = 0$, obtained by variation of the classical solution, 
$\m (r; r_1, r_2)$ with respect to the parameters $(r_1, r_2)$. Indeed by changing variables from $r$ to $w$ and using 
the explicit solution (\ref{hshell}) and (\ref{req}) it is readily verified that one solution to $\cL\chi_{_0}=0$ is $\chi_{_0}=1-w$, 
from which the second linearly independent solution $(1-w)\ln\,w + 4$ may be obtained. Since these correspond to 
varying the positions of the $r_1,r_2$ interfaces, neither $\chi_{_0}$ vanishes at $(r_1, r_2)$ and neither is a true zero 
mode.  However, we may set $\del \m = \chi_{_0}\, \psi$, where $\psi$ does vanish at the endpoints and insert this into the 
second variation (\ref{varent}). Integrating by parts, using the vanishing of $\del \m$ at the endpoints and $\cL \chi_{_0} = 0$ 
one obtains
\vspace{-4mm}
\be
\del^2 S= -\frac{a k_B}{4\hbar \GN}\int_{r_1}^{r_2}\!r\, dr\, \left(\frac{d\m}{dr}\right)^{\!\!-\frac{3}{2}}
h^{-\frac{1}{2}}\chi_{_0}^2\,\left(\frac{d\psi}{dr}\right)^{\!2} < 0
\label{secvar}
\vspace{-4mm}
\ee
which is negative~definite.

Thus, the~entropy of the solution is maximized with respect to radial variations that vanish at the endpoints, 
i.e.~those with fixed total energy. Since deformations with non-zero angular momentum decrease the 
entropy even further, stability under radial variations is sufficient to demonstrate that the solution is 
stable to all small perturbations. In~the context of a hydrodynamic treatment, thermodynamic stability 
is also a necessary and sufficient condition for the {\it dynamical} stability of a static, spherically symmetric 
solution of Einstein's equations~\cite{Cocke:1965}.

\section{Conclusions} 
\label{Sec:Concl}

A compact, non-singular solution of Einstein's equations has been presented here as a possible stable alternative 
to BHs for the endpoint of gravitational collapse~\cite{gravastar:2001}. Realizing this alternative requires that a quantum 
gravitational vacuum phase transition intervene and allow the vacuum energy $\rho_{_V} = -p_{_V}$ to change
before the classical event horizon or a trapped surface can form. Although~only the static spherically symmetric case has been 
considered, it is clear on physical grounds that axisymmetric rotating solutions should exist as well. Since the entropy 
of these objects is of the order of magnitude of a typical stellar progenitor, or~less,  there is no huge BH entropy to be explained
and instead a process of entropy shedding, as~in a supernova, is needed to produce a cold GBEC or `grava(c)star'~remnant. 

In this paper we have assumed that the thin boundary layer where the quantum phase transition occurs
can be described as a relativistic fluid with maximally stiff equation of state $p= +\rho$, where the speed of light 
is equal to the speed of sound. Although~this is a phenomenological model, the~possibility that such a 
boundary layer could be expected to produce excitations bearing the imprint of its fundamental normal
mode vibration frequencies when struck should be robust and serve to distinguish gravastars from black 
holes observationally. These surface excitations may also provide a more efficient central engine for 
astrophysical sources to impart energy to accreting matter, producing ultra-high energy particles, gamma 
rays and gravitational radiation. Finally, the~interior dS region with $p_{_V} = -\rho_{_V}$ may be 
interpreted also as a cosmological spacetime, with~the horizon of the expanding universe replaced by 
a quantum phase~interface.



\bibliographystyle{apsrev4-1}
\bibliography{gravity21Aug}

\appendix
\section{Gravitational Condensate Stars: Further Developments}
The main text of this paper is a minimally corrected version of the previously unpublished arXiv submission~\cite{gravastar:2001},
in~which the original proposal that the final state of complete gravitational collapse is a non-singular gravitational vacuum 
condensate star (`gravastar') was made. A~somewhat expanded version of this paper appeared in~\cite{MazEMPNAS:2004}. 
The authors take this opportunity to provide an extended Appendix, updating the status of the gravastar proposal, collecting 
under seven subtitles the most significant developments over the past two decades relating to this proposal, with additional
explanation and annotations for each of the following appendices.

\subsection{Background: Preliminary Description the Boundary Layer}
\label{Sec:Background}

Discussions of matching the exterior Schwarzschild exterior solution to a non-singular de Sitter (dS) interior had a long
 history. Continuous transitions between the two were studied, e.g.~in~\cite{Dymn:1992}, while it was
recognized that joining the exact Schwarzschild and dS geometries directly at their mutual horizons $H^{-1}$ 
and $2GM$, requires some discontinuity or interposition of `non-inflationary material'~\cite{PoisIsr:1988}. In~addition 
to uncertainties of the physics involved, the~earlier GR formalism~\cite{Lanczos:1924,OBrSyn:1952,Israel:1966a,Israel:1967}
for dealing with singular hypersurfaces when the normal to hypersurface becomes null, as~it does at a BH horizon, were 
recognized to be inadequate~\cite{BarIsr:1991}. The~necessity of some anisotropic matter at the joining of the
interior to exterior geometries was made explicit in~\cite{CatVis:2005}. 

Partly for the reason of avoiding the technical difficulties associated with singular null hypersurfaces, the~proposal in the
original gravastar paper~\cite{gravastar:2001} here makes use of two timelike hypersurfaces at $r_1$ and $r_2$ with an 
interposed fluid boundary layer of `non-inflationary material' obeying the  equation of state $p=\rho$. The~choice of this equation of state 
at the causal limit where the speed of sound coincides with the speed of light, was motivated by physical considerations of 
a quantum phase transition produced by the infrared effects of dimensional reduction from $D=4$ to $D=2$ dimensions.

This choice was also motivated in part by the observation of `t Hooft that a self-screening Hawking atmosphere of a fluid 
with $p=\ka \rho$ near to the horizon could produce the $1/4$ area law of the Bekenstein--Hawking entropy $S_{BH}$
when $\ka =1$~\cite{tHooft:1998}, however at the price of an interior negative mass singularity, suggesting a repulsive core.
Further physical considerations of quantum phase transitions in condensed matter analogs in~\cite{ChaplineHCL:2001} also suggested 
that an equation of state at the extreme causal limit should play a role. Nevertheless, the~choice of $p=\rho$ in~\cite{gravastar:2001} 
is certainly a phenomenological ansatz, illustrating a proof of principle, but~without a rigorous basis in fundamental physics. 
It is therefore subject to modification as that fundamental physics came more clearly into view by subsequent developments~\cite{EMVau:2006,EMZak:2010,EMEFT:2022}.

\subsection{The Macroscopic Effects of the Conformal Anomaly and Value of \texorpdfstring{$\ve$}{}}
\label{Sec:Anom}

A major step in providing a rigorous basis from quantum theory of large effects at horizons came in 2006, with~the observation 
that the energy--momentum tensor derived from the effective action of the conformal trace anomaly of massless fields in curved
space becomes large (indeed formally infinite) for generic quantum states at both the Schwarzschild BH and dS static
horizons~\cite{EMVau:2006}. The~conformal anomaly becomes relevant at horizons because of the conformal behavior of the 
near horizon geometry, typified by the extreme blueshifting of local frequencies and energies there, making all finite mass 
scales negligible as $r\to \rM$ from outside, or~$r\to \rH$ from inside~\cite{EMZak:2010,AntMazEM:2012}. The~effective 
action of the conformal anomaly and $T^\m_{\ \n}$ derived from it provides a clear basis in quantum field theory (QFT)
for large semi-classical backreaction effects on BH and dS horizons~\cite{GiaEM:2009,EMSGW:2017}, consistent with general 
covariance and the weak equivalence~principle.

These semi-classical vacuum polarization effects occur in zero temperature QFT and, thus obviate any need to invoke
an ultra-relativistic fluid ansatz with the $p=\rho$ equation of state, or~any `fluid' temperature at all as in (\ref{localent}) in the thin 
shell region. A~gravastar relying on quantum vacuum polarization can be at precisely {\it zero temperature} and a true quantum
endpoint for complete gravitational collapse. The~implication that the GBEC must be at a very low (if not identically zero temperature) 
was inherent in the original arXiv article~\cite{gravastar:2001}, and~presented in seminars at the time, including at the Inst.~for Theoretical Physics
(Univ.~Calif.~Santa Barbara) on 9 May 2002~\cite{EM_ITP:2002}. Although~the gravastar surface at $r=\rM$ is a `wall' of large 
vacuum stresses, in~some ways similar to the `firewall' later discussed in~\cite{AMPS:2013}, the~boundary layer of a GBEC
is at low temperature, and hence not a `firewall.' Nor does it imply or require a catastrophic breakdown of semi-classical gravity
or causality in order to eliminate the various BH information paradoxes~\cite{MottVauPT}. 

The anomaly stress tensor $T^\m_{\ \n}$ of the mean field description of semi-classical gravity becomes large enough to affect 
the classical geometry when $\D r \sim L_{\rm Pl}$. This determines the small parameter $\ve$ of the main text and original~\cite{gravastar:2001} 
to be of order $L_{\rm Pl}/\rM$ (or $L_{\rm Pl}/\rH$), where the quantum effects of the anomaly must be taken into account. 
The {\it physical} proper thickness of the thin shell is,therefore~\cite{EMZak:2010}
\vspace{-2mm}
\be
\ell \sim \sqrt {L_{\rm Pl}\, \rM} = 2.2\times 10^{-14} \sqrt{M/M_\odot}\  \, {\rm cm}.
\label{ellvalue}
\vspace{-2mm}
\ee
It is significant that $L_{\rm Pl} \ll \ell \ll \rM$ so that the shell is very thin on macroscopic scales, making it
challenging to detect in astronomical observations, but~nevertheless very much larger than the microscopic Planck scale at which 
the semi-classical approximation breaks down. For~a solar mass QBH $\ve \sim 10^{-38}$ \cite{MazEMPNAS:2004}, which well 
justifies the $\ve \ll 1$ approximation~(\ref{hshell}) of the~text.

\subsection{The Schwarzschild Interior Solution and Determination of \texorpdfstring{$C$}{}}
\label{Sec:C}

An independent but equally significant development came in 2015 with the realization that an infinitely thin shell gravastar solution 
to Einstein's equations actually results from the 1916 interior solution of a constant density star by Schwarzschild~\cite{Schwarz:1916}, 
provided the limit is taken in which the surface of the star $r_{\rm star}$ is at the horizon $\rM$ itself~\cite{MazEM:2015}. Remarkably, 
in that limit, the~1916 Schwarzschild constant density solution produces a $p=-\rho$ interior, which is just the gravitational condensate 
star of the main text and original~\cite{gravastar:2001}. This dS interior `fluid' has no ordinary sound modes, and, therefore, removes 
Einstein's original objection to the Schwarzschild constant density interior, in~this limiting~case. 

This limiting case of the Schwarzschild interior solution also unambiguously determines the constant $C$ which was
undetermined in~\cite{gravastar:2001} to be~\cite{MazEM:2015}
\vspace{-3mm}
\be
C = \sdfrac{1}{4}
\label{Cvalue}
\vspace{-2mm}
\ee
by the matching of the interior dS time to the exterior Schwarzschild time. This turns out to
be exactly the value necessary to make the surface gravity
\vspace{-3mm}
\be
\ka = \sdfrac{1}{2} \sqrt{\sdfrac{h}{f}} \sdfrac{df}{dr}
\label{surfgrav}
\vspace{-2mm}
\ee
of the two geometries, $\ka_-$ and $\ka_+$, to~be {\it equal and opposite}, a~necessary condition for the forces 
on each side of the membrane to balance (and the corresponding periodicities and Hawking temperatures 
to be equal in the Euclidean continuation). The~discontinuity in the surface gravities
\vspace{-2mm}
\be
\D \ka = \ka_+ - \ka_- = 2 \ka_+ = \frac{1}{4 GM}> 0
\label{Delkap}
\vspace{-2mm}
\ee
also unambiguously determines the physical surface tension $\tau_S = \D\ka/8 \p G$ of the membrane boundary.
The presence of the $\del$-function in the transverse pressure $p_\perp \neq p$ at the membrane interface this implies
provides the loophole in the Buchdahl bound, which had assumed isotropic fluid pressure $p_\perp =p$ throughout~\cite{Buchdahl:1959},
and is consistent with the general results of~\cite{PoisIsr:1988,CatVis:2005} requiring `non-inflationary' anisotropic stresses at the~joining. The surface tension of the membrane also shows that the First Law of spherical~gravastars 
\vspace{-4mm}
\be
dM = dE_{_V} + \tau_S\, dA
\vspace{-3mm}
\ee
expressing energy conservation, is a {\it purely mechanical classical} relation at {\it zero} temperature and zero entropy, 
in which neither $\hbar$ nor $k_B$ appear. This makes sense of the original BH Smarr relation~\cite{Smarr:1973}, by~correcting it
by the factor of $2$ in (\ref{Delkap}) to account for the {\it difference} in surface gravities, rather than just $\ka_+$. The~positivity 
of $\tau_S$ further shows that deforming the surface by increasing the area requires energy, indicating its stability to perturbations, 
without reliance on the thermodynamic stability argument of Section~\ref{Sec:Stability}~\cite{PosChir:2019,ChirPosGued:2020}. Note that both 
the value and the sign of $\tau_S$ obtained in the classical fully GR analysis of~\cite{MazEM:2015} differ from what was conjectured 
in the flat space condensed matter analog model of~\cite{ChaplineHCL:2001}.

An additional serendipitous consequence of this reanalysis of the 1916 interior solution is that it provides an explicit example 
of `gluing' of two different geometries at their mutual null horizons, in~which the surface stresses can be unambiguously
determined, providing a clear interpretation of the surface tension of the null surface at $r=\rM$. This shows that gluing 
the exterior Schwarschild to interior dS geometries {\it directly} is indeed possible in classical GR, with~(\ref{Cvalue}), 
and~has served 
to provide the general matching conditions for null surfaces with non-zero angular momentum as well~\cite{BelGonEM1:2022}.
This improved understanding and generalization of the junction conditions to null hypersurfaces and, in~particular, 
rotating null horizons appropriate for the Kerr geometry opens the way to finding rotating gravastar solutions,
the study of which in the case of slow rotation following methods of~\cite{Hartle:1967,ChandraMiller:1974} has 
begun~\cite{Maztalk:2020,Posada:2017,BelGonEM2:2022}. These results already indicate that the moment of inertia $I$
of a slowly rotating gravastar, defined as the ratio of its angular momentum $J$ to angular velocity $\omega_{_H}$
of its thin shell located at the Kerr horizon, is
\vspace{-2mm}
\be
I = \frac{J}{\omega_{_H}\!\!} = M \rM^2 = 4G^2 M^3
\label{MomIner}
\vspace{-2mm}
\ee
consistent with the BH ``no hair'' theorems being extended to rotating gravastars as
well~\cite{Maztalk:2020,BarCarRubLib:2019}, 
at least in the strictly classical limit of $\ve \to 0$. Gravastar ``hair'' for a surface layer of finite thickness $\ell$ (\ref{ellvalue})
would be limited to that quantum phase transition boundary layer of a thin shell thickness (\ref{ellvalue}) only.

For the original spherically symmetric gravastar solution, the~proper matching at the null horizon and the condition (\ref{Cvalue}) 
completely eliminates the two independent spacelike boundaries at $r_1$ and $r_2$ and intermediate region II $p=\rho$ layer 
(\ref{match})--(\ref{r1r2C}). This is the {\it universal} thin shell limit of a non-rotating gravastar in the classical limit 
$\ve \to 0$, in~the sense of being independent of any assumptions of an equation of state of the surface layer or any other 
matching conditions. The~solution of~\cite{MazEM:2015} has a surface layer of infinitesimally small thickness, with~a stress tensor 
that is precisely a Dirac $\del$-function on the horizon in this~limit. 

\subsection{Thin Shell vs. Thick Shell}
\label{Sec:Shell}

The distinguishing feature of the original gravitational condensate star proposal of~\cite{gravastar:2001} of the main text is the
abrupt change in ground state vacuum energy at the horizon, characteristic of a quantum phase transition there. This should be 
clear from the essential role of the horizon as a infinite red shift surface in both~\cite{ChaplineHCL:2001,gravastar:2001}, 
the assumption of $\ve \ll 1$ and the estimate of $\ve$ and $\ell$ in Appendix \ref{Sec:Anom} from the conformal anomaly. The~proper length 
$\ell$ determined by the stress tensor of the conformal anomaly takes the place of the `healing length' introduced, but~left undetermined 
in the analogy of the horizon in GR to the non-relativistic quantum critical surface of a sound horizon in~\cite{ChaplineHCL:2001}. Thus, 
the term `gravastar' should apply only to the gravitational condensate star model of~\cite{gravastar:2001} in the text, described 
also in~\cite{MazEMPNAS:2004}, and~further refined in~\cite{MazEM:2015}, where the lightlike null horizon clearly plays a privileged role 
as the locus of joining of interior and exterior classical geometries, with~equal and opposite surface gravities, and~where the conformal
anomaly stress tensor also grows large, and~a quantum phase transition can~occur.

Despite this physically privileged role of the horizon in the original gravastar proposal~\cite{gravastar:2001}, a~number of papers 
appeared subsequently that discussed what may be called `generalized gravastars', or regular solutions with macroscopically 
large or `thick' shells, comparable to the gravitational radius $\rM$ itself, with~compactness $GM/r$ differing from the maximal value of $1/2$, 
by order unity, some time varying, or~with timelike surfaces displaced from the Schwarzschild or dS horizons by finite amounts~\cite{VisserWilt:2003,DymnGalak:2005,Lobo:2006,DeBenHorvat:2006,ChirRez:2007,HorvIlij:2007,BertiCardPani:2009,BertiCardPani:2010,
Mar-MorGarLoboVisser:2012,SakaiGravShadow:2014,Pani:2015,Uchikata:2015,ChirRez:2016,UchiPani:2016,VolkKok:2017,RaySenNim:2020,SenGhoRayMisTri:2020}.
Several authors proceeded to discuss both ergoregion instabilities and observational bounds on such hypothetical objects, with~various 
assumptions about boundary conditions and surface matchings~\cite{ChirRez:2008,CardPaniCadCavPRD:2008, CardPaniCadCavCQG:2008}.
It should be clear that these instabilities or observational bounds do not apply to gravastars, which {\it by definition} are static 
configurations with an {\it infinitesimally thin shell located at the horizon}, for~$\ve =0$, or~straddling and replacing the would-be 
classical Schwarzschild and dS horizons for very small finite $\ve$, with~metric functions $f \sim h = \cO(\ve)$ there. Any other regular QBH
is not a~gravastar.

\subsection{The Status of Constraints from Astronomical~Observations}
\label{Sec:Astro}

Because the exterior geometry of a gravastar is identical to that of a classical BH down to the scales of its very thin shell surface
boundary layer at $\ell$ given by (\ref{ellvalue}) above the would-be classical horizon, it should be clear that a gravastar will be cold, 
dark and indistinguishable from a BH by almost all traditional astronomical observations. Any radiation from such a deeply redshifted 
surface can escape to infinity only if emitted from a tiny 'pinhole' solid angle less than of order $\ve$ from the perpendicular, or~it will 
fall back onto the surface. Attempts to `prove' the existence of a BH horizon or absence of a surface from the absence of thermal radiation 
and/or absence of X-ray bursts which would be expected {\it if} the surface is composed of conventional matter, and~{\it if} any advected 
matter deposited onto the surface is re-radiated rather than absorbed, are therefore bound to fail. This point was succinctly made 
in~\cite{AbramKluzLas:2002}, soon after the gravastar proposal of~\cite{gravastar:2001}. 

The~authors of~\cite{AbramKluzLas:2002}
also recognized that any surface of an ultracompact QBH was bound {\it not} to be composed of conventional matter, such as a 
neutron star crust, needed for the thermonuclear reactions that give rise to X-ray bursts. Moreover, in~order for the gravastar 
proposal to be a viable alternative for a BH of {\it any} mass, a~gravastar must be able to absorb accreting baryonic matter and 
convert it to the interior condensate, thereby growing its mass to any larger value. Any substantial efficiency of absorption 
and conversion of energy to interior condensate would reduce the energy re-radiated and make the object dark in most if not 
all the observable electromagnetic~spectrum. 

The authors of~\cite{BrodNary:2007} argued for quite stringent limits on what they called `gravastar' models, assuming thermalization 
of accreting  matter in a steady state emission. Aside from: (i) an unjustified and rather {\it ad~hoc} assumption of the form for internal 
energy and heat capacity of the `matter' supposed to be composing the QBH, (ii) not accounting for the relativistic pinhole effect 
suppressing all emission from a deeply redshifted surface, and~(iii) ignoring the possibility of near total absorption of accreting matter 
without any heating of the QBH, which would all but eliminate any thermal re-emission with sufficiently gentle accretion, the~arguments 
of~\cite{BrodNary:2007} were attempts to constrain the condensed matter analog model of~\cite{Chap:2003,Chap:2005}. This in~any 
case, is not the gravastar described in~\cite{gravastar:2001,MazEMPNAS:2004}, this article, or~the later~\cite{MazEM:2015}.  

Similar arguments based on thermalization and steady state re-emission of radiation, again ignoring the possibility of absorption by the 
QBH surface, with~claims of strong observational bounds were made in~\cite{NarayMc:2008,BrodNarayDoel_ApJ:2015}. These
unjustifiably strong claims of `proof' of BH horizons and the assumptions upon which they are based have been critically examined 
by several authors~\cite{CardPani:2017,CarbDiFiLibViss:2018,CardPani:2019,CarbRubDiFilLibVis:2022}, and~shown to be flawed. 
These authors showed first that the
assumption that thermodynamic equilibrium can be established between an accretion disk and the QBH on a 
reasonably short timescale is incorrect for a deeply redshifted surface for $\ve \to 0$, due to the gravitational lensing pinhole effect,
already pointed out in~\cite{AbramKluzLas:2002}. The~best limits one can obtain from the observations of M87 or Sag A* when
this classical GR effect is taken into account is in the range of $\ve < 10^{-15}\ {\rm to}\ 10^{-17}$~\cite{CarbRubDiFilLibVis:2022}, 
impressive, but~still many orders of magnitude short of $10^{-38}$ expected for a gravastar. Secondly, the~energy emitted was assumed 
to be electromagnetic in observable wavelengths, whereas a sizable fraction of any re-emitted energy could be in the form of neutrinos 
or in unobserved radiation~\cite{CardPani:2017,CardPani:2019}. Thirdly, and~most importantly, a~sizable fraction even approaching 
unity of the accreting matter may be {\it absorbed} by the gravastar, with~virtual no re-emission whatsoever. As~a result, there are no 
useful bounds from the non-observation of electromagnetic emission from any astrophysical QBH, and~the possibility that they may 
all be gravastars with $\ve \ll 10^{-17}$ remains~open.

The converse claim of a {\it lower} bound of $\ve \gtrsim 10^{-24}$ in~\cite{CarbKumarLu:2018} is based on a strong assumption 
of the restrictive form of the Vaidya metric and stress tensor in the vicinity of the QBH surface, setting to zero all of its components 
except $T_{vv}$ in advanced null coordinates. This bound also disappears if the assumption upon which it is based is relaxed,
which it almost certainly should~be.

\subsection{Gravitational Waves and~Echoes}
\label{Sec:GW}

The observation of gravitational waves (GWs) by LIGO~\cite{LIGO1:2016} has opened up a new window on the universe that among 
many other interesting possibilities provides perhaps the best opportunity for observational test of the gravastar proposal. The~GW data are 
not yet accurate enough to test the prediction of a {\it discrete} spectrum of ringdown modes from a non-singular gravastar with a surface 
made in~\cite{MazEM:2015}. Indeed it was quickly realized that sensitivity to the nature of a very compact QBH with $\ve \ll 1$ is obtained 
only with some delay time after the initial GW merger signal, in~the ringdown phase~\cite{CardFranPani:2016}, where the signal/noise ratio is 
very much lower. Nevertheless a regular QBH such as a gravastar could produce a GW `echo' at multiples of the characteristic time 
\vspace{-3mm}
\be
\D t \sim 2GM\, \ln (1/\ve)
\label{Delt}
\vspace{-2mm}
\ee
after the compact object merger event~\cite{CardHopMacPalPani:2016}. These may be observable with the improved sensitivities of
Advanced LIGO, Virgo, and~future~detectors. 

The basis for such echoes is the expectation that GWs produced in the merger could reflect from the internal centrifugal barrier 
of a gravastar and re-emerge with a logarithmically long time delay for $\ve \ll 1$, thus, in~principle, opening up the possibility of 
testing GR and the nature of QBH's on scales very close to the would-be horizon. 

A~somewhat different scenario was considered 
in~\cite{AbediDykAfsh:2017}, with~a claim of tentative evidence for an echo signal in the LIGO/LSC data~\cite{QBHsPI:2017}.
However, an~analysis of the same data by members of the LIGO/LSC collaboration concluded that the echo signal was just $1.5 \s$ 
above the noise level~\cite{WesterNiel:2018}. The~subject of GW echoes from QBH's such as gravastars continues as an area 
of active research~\cite{BarCarGar:2017,VolkKok:2017,MasVolkKok:2017,WangOshAfs:2020}, requiring substantially more data 
from Advanced LIGO, Virgo, and~successor detectors to settle this question~\cite{CardPani:2019}. 

\subsection{The EFT of Gravity and Dynamical Vacuum~Energy}
\label{Sec:EFT}

A complete dynamical model of gravitational condensate stars has been lacking for the two decades since~\cite{gravastar:2001},
and, in~particular, the~mechanism by which $p_{_V}=-\rho_{_V}$ vacuum energy can change at a would-be BH horizon. In~this past year,
just such an effective field theory (EFT) of gravity in which $\La_{\rm eff}$ is described by a dynamical four-form gauge field coupled
to the Euler--Gauss--Bonnet term of the conformal anomaly in the presence of torsion has been proposed~\cite{EMEFT:2022}.
This EFT promises to provide the theoretical Lagrangian basis for development of the gravastar proposal first made in 2001, 
by explicit gravastar solutions in which the classical, coherent four-form field strength is the explicit realization of the
gravitational condensate hypothesized in~\cite{gravastar:2001,MazEMPNAS:2004} and the text. In~this fully dynamical EFT 
of vacuum energy, $\La_{\rm eff}$ couples to the conformalon field of the conformal anomaly, and~both change rapidly 
near the horizon worldtube  of $\mathbb{R} \times \mathbb{S}^2$ topology. 

The EFT of~\cite{EMEFT:2022} provides the Euler--Lagrange equations which should exhibit a static, gravastar solution.
Linearization about this solution will then enable a definitive study of the dynamical stability of the gravastar and determine 
its normal modes and frequencies of vibration, relevant for GW observations. This EFT also paves the way for studying rapidly 
rotating gravastars and their dynamical collapse formation process as well, and~is expected to provide the basis for quantitative 
predictions to be compared to the increasing amount of GW and other astrophysical data expected to be provide by aLIGO, VIRGO, and~
other observations in the next two~decades.

\end{document}